\documentclass[12pt]{iopart}

\usepackage{iopams}
\usepackage{graphicx}
\usepackage{color}

\begin{document}

\title[Non-adiabatic topological spin pumping]
{Non-adiabatic topological spin pumping}

\author{W Y Deng$^1$, W Luo$^1$, H Geng$^1$, M N Chen$^1$, L Sheng$^{1,2}$ and D Y Xing$^{1,2}$}

\address{
$^1$ National Laboratory of Solid State Microstructures and Department of Physics, Nanjing University, Nanjing 210093, China\\
$^2$ Collaborative Innovation Center of Advanced Microstructures,
Nanjing University, Nanjing 210093, China

E-mail: shengli@nju.edu.cn and dyxing@nju.edu.cn}

\begin{abstract}
Based on the Floquet scattering theory, we analytically investigate
the topological spin pumping for an exactly solvable model.
Floquet spin Chern numbers are introduced to characterize the periodically time-dependent
system. The topological spin pumping remains
robust both in the presence and in the absence of the
time-reversal symmetry, as long as the pumping frequency
is smaller than the band gap, where the electron transport involves only
the Floquet evanescent modes in the pump. For the pumping frequency
greater than the band gap, where the
propagating modes in the pump participate in the electron transport,
the spin pumping rate decays rapidly,
marking the end of the topological pumping regime.
\end{abstract}


\maketitle
\section{Introduction}\label{s1}
Generally speaking, quantum pumping is a dynamic transport mechanism that dc charge currents can flow under zero bias via a quantum system, in which some parameters are periodically modulated in time. It was originally proposed by Thouless and co-workers in the 1980s, who found
that a quantized charge can be pumped during a period of slow variation of potential in the Schr\"{o}dinger equation~\cite{Thouless,Niu}.
The amount of charge pumped per cycle is directly related to a topological invariant of the
system, namely, the Chern number~\cite{Thouless,Niu}. The same topological invariant was also used to
classify the integer quantum Hall effect~\cite{TKNN}. There have been continuous interest in the topological charge pumping~\cite{Zhou,Qi1,Mahfouzi,Ueda,Wang,Gibertini,Marra}.

For weak pumping, where the time-dependent parameters vary so slowly that the system can be treated adiabatically, the pumped current can be obtained by the Berry phase theorem associated with the instant scattering matrix~\cite{Buttiker,Brouwer}.
As has been clearly shown by Graf and coworkers~\cite{Graf,Braunlich},
the topological approach and the scattering matrix one provide equivalent descriptions of
the adiabatic quantum pumping. However, when the charge pumping was observed in an open quantum dot experimentally, Switkes $et$ $al.$ found
that the adiabatic theory is inadequate to explain the observation in the
strong pumping regime~\cite{Swithes}. Then Zhu and Wang developed the Floquet scattering method to study the pumping with a series of time-periodic potentials, their main results in both the weak pumping and strong pumping regimes being consistent with experiment results~\cite{Zhu}. Moskalets and B\"{u}ttiker generalized the Floquet scattering theory to the quantum pumping in mesoscopic conductors~\cite{Moskalets}. Kim found that the Floquet scattering approach and the adiabatic scattering approach give the exactly equivalent result under the weak pumping condition for a double delta-barriers model~\cite{Kim}. The essential difference of the two methods is that the adiabatic condition is necessary for the adiabatic pumping, while the Floquet pumping only relies on the periodicity of the time-dependent parameters. Moreover, in the Floquet theory, the pumped current may be nonzero when the phase difference between the driven potentials is zero~\cite{Zhu} or only a single parameter is time-dependent in the system~\cite{Prada}, although the adiabatic theory predicts vanishing pumped current under these conditions.

The topological spin pump is a spin analogue of the Thouless charge pump,
which was proposed after the discovery of the topological
insulators~\cite{Kane,Bernevig,Konig}. Topological insulators have nontrivial bulk band topology
characterized by unconventional topological invariants~\cite{Hasan,Qi}. Two-dimensional topological insulators are also called the quantum spin Hall (QSH) systems, whose
topological properties are usually described by the ${Z_2}$ invariant~\cite{Kane1} or spin Chern numbers~\cite{Sheng,Prodan}. An important consequence of the nontrivial
bulk band topology of the QSH systems is that a pair of helical edge states
emerge in the band gap, which make them conductive at the sample boundary
and essentially different from ordinary insulators. The ${Z_2}$ invariant is
well-defined only when the time-reversal (TR) symmetry is present~\cite{Kane1}. This feature is consistent with the fact that the edge states
in the QSH systems are gapless in the presence of the TR symmetry, and usually gapped otherwise.
While the spin Chern numbers yield an equivalent description
for TR-invariant systems,
their robustness does not rely on any symmetries~\cite{Prodan,Li,Yang}. Nonzero spin Chern numbers
guarantee that edge states emerge in the bulk band gap, which could be gapless or
gapped, depending on the symmetry and local microscopic structures of the sample
edges~\cite{Sheng1}.

The topological spin pump has an intimate connection to
the topological invariants underlying the QSH effect. As an
observable effect, it provides a possible route to
investigate the topological invariants experimentally.
Based on a spin-conserved model of an antiferromagnetic chain,
Shindou proposed the prototype of a topological spin pump~\cite{Shindou}.
Fu and Kane established the more general concept of a ${Z_2}$ pump
without limitation of spin conservation~\cite{Fu}. In the $Z_{2}$ pump,
while the amount of spin pumped
per cycle is not integer-quantized
in the absence of spin conservation, the pumping process is protected
by a $Z_2$ topological invariant, provided that
the TR symmetry is present. Meidan, Micklitz, and Brouwer
classified topological spin pumps
based on general properties of the scattering matrix~\cite{Meidan}.
They showed that in the weak coupling limit, topological spin
pumps are characterized by the appearance of symmetry-protected gapless end states
during the pumping cycle, similarly to the $Z_2$ pump. Several
different methods have been put forward to realize the $Z_2$ pump experimentally,
including a Luttinger liquid~\cite{Sharma}, a double-corner junction
in a topological insulator~\cite{Citro}, and quantum wires proximity coupled to a
superconductor~\cite{Keselman}.

Recently, Zhou $et$ $al.$~\cite{Zhou2} investigated the effect of TR
symmetry breaking on the topological spin pumping by introducing randomly distributed
magnetic impurities with classical spins
into the one-dimensional model used by Fu and Kane~\cite{Fu}.
In contradiction to the previous belief,
the magnetic impurities only affect the amount of spin pumped per cycle in a perturbative manner
rather than destroy the spin pumping effect immediately.
While the $Z_2$ invariant can no longer be defined in this situation
as the TR symmetry is explicitly broken,
the spin pumping effect can be attributed to the spectral flow of the spin-polarized
Wannier functions driven by nonzero spin
Chern numbers. Chen $et$ $al.$ proposed that such spin-Chern pumping effect might be
observed in a two-dimensional topological insulator with oscillating dual gate voltages
subject to an in-plane $ac$ electric field~\cite{Chen}. In the absence of disorder, they showed
that the effective one-dimensional system for each transverse momentum between two
critical values $-k^{c}_{y}$ and $k^{c}_{y}$ acts as a
spin-Chern pump, and all the individual transverse momenta between $-k^{c}_{y}$ and $k^{c}_{y}$
join together to contribute to a bulk spin pumping
current in proportion to the width of the pump.

So far, existing theoretical works on the topological spin pump
 are limited to the adiabatic regime, which is difficult to reach
in experiments. It is unclear how the topological properties of the spin pump evolve and
whether the topological spin pumping remains to be robust in the more general non-adiabatic
regime. Therefore, theoretical study of the topological spin pumping
beyond the adiabatic approximation is highly desirable.
In this work, we study the non-adiabatic topological spin pumping based on an
exactly solvable model. We introduce
the Floquet spin Chern numbers to describe
the topological properties of the spin pump, which
extend the adiabatic spin Chern numbers
to more general non-adiabatic regime. Based on the Floquet scattering matrix theory,
the spin pumped per cycle is calculated from
the Floquet scattering matrix theory in the absence and
in the presence of the TR symmetry. We show that
spin pumping through the Floquet evanescent modes remains
robust and insensitive to the parameters of the system, as long as the pumping frequency
is smaller than the band gap. For the pumping frequency greater than the band gap,
the spin pumping involves the propagating modes in the pump, and decreases rapidly,
which marks the boundary of the topological spin pumping regime.

In the next section, we first introduce the model Hamiltonian. Then we define the Floquet spin Chern numbers, which are applicable to the more general non-adiabatic regime. In section~\ref{s3}, the spin pumped per cycle in the presence of the TR symmetry is calculated by using the Floquet scattering matrix method. In section~\ref{s4}, the spin pumped in the absence of the TR symmetry is calculated, and
the result is compared with that obtained in the adiabatic approximation. The final
section contains a summary.

\section{Floquet spin Chern numbers}\label{s2}
Let us start from the effective continuum electron model of a one-dimensional
zigzag atomic chain proposed in~\cite{Zhou2}, which was used to describe the body of a topological spin pump
\begin{equation}\label{eq1}
{H_P} = {v_{\rm{F}}}{p_x}{\sigma _y} + \alpha \left( t \right){\sigma _x} + g\left( t \right){s_z}{\sigma _z}\ .
\end{equation}
Here, $s_{z}$ is the Pauli matrix describing electron spin, and $\sigma_{x(y,z)}$ are the Pauli matrices associated with sublattices. The first term originates from the nearest neighbor hopping
with ${v_{\rm{F}}}$ as the Fermi velocity. The second term can be induced by agitating oscillatory shear deformation of the substrate, on which the one-dimensional atomic chain is deposited,
with $\alpha \left( t \right) = {\alpha _0}\cos \omega t$ and ${\alpha _0}$ as the strength of the deformation potential. The last term is the time-dependent Zeeman splitting energy
alternating on the AB sublattices with $g\left( t \right) = {g_0}\sin \omega t$
and $g_{0}$ as the strength of the Zeeman field.

The wavefunctions of the time-dependent Hamiltonian Eq.\ (1) is difficult to
obtain for arbitrary parameter sets. However, we find
that in the special case ${\alpha _0} = {g_0}$, the problem
is exactly solvable. We will confine ourselves to this case, but the conclusion is
expected to be applicable to more general cases. The Hamiltonian ${H_P}$ conserves
spin ${s_z}$, so that the two spin species can be treated separately.
The time-dependent Schr\"{o}dinger equation for the spin-up electrons is
\begin{equation}\label{eq2}
i\hbar \frac{{\partial {\Psi _{P, \uparrow }}\left( {x,t} \right)}}{{\partial t}} = {H_{P, \uparrow }}{\Psi _{P, \uparrow }}\left( {x,t} \right)\ ,
\end{equation}
\begin{equation}\label{eq3}
{H_{P, \uparrow }} = {v_{\rm{F}}}{p_x}{\sigma _y} + \alpha \left( t \right){\sigma _x} + g\left( t \right){\sigma _z}\ .
\end{equation}
Since the Hamiltonian is periodic in time, the Floquet theorem is applicable to
this equation. By setting ${\Psi _{P, \uparrow }}\left( {x,t} \right) = {e^{ - i{E_{Fl}}t/\hbar }}{\psi _l}\left( {x,t} \right)$ , where ${E_{Fl}}\in [0,\hbar\omega)$ is the Floquet eigenenergy and ${\psi _l}\left( {x,t} \right)$ is a periodic function in time ${\psi _l}\left( {x,t} \right) = {\psi _l}\left( {x,t + T} \right)$  with period $T = 2\pi /\omega $, equations (2) and (3) are derived to be
\begin{equation}\label{eq4}
i\hbar \frac{{\partial {\psi _l}\left( {x,t} \right)}}{{\partial t}} = H_{P, \uparrow }^l{\psi _l}\left( {x,t} \right)\ ,
\end{equation}
\begin{equation}\label{eq5}
H_{P, \uparrow }^l = {v_{\rm{F}}}{p_x}{\sigma _y} + \alpha \left( t \right){\sigma _x} + g\left( t \right){\sigma _z} - {E_{Fl}}\ .
\end{equation}

Using a unitary transformation ${\psi _r}\left( {x,t} \right) = U{\psi _l}\left( {x,t} \right)$
with $U = {e^{ - i\omega t{\sigma _y}/2}}$ , which means transforming
the system from the laboratory reference frame to a rotating reference frame, we obtain
\begin{equation}\label{eq6}
i\hbar \frac{{\partial {\psi _r}\left( {x,t} \right)}}{{\partial t}} = {H_r}{\psi _r}\left( {x,t} \right),
\end{equation}
\begin{eqnarray}\label{eq7}
\nonumber
{H_r} &=& i\hbar \frac{{\partial U}}{{\partial t}}{U^\dag } + UH_{P, \uparrow }^l\left( t \right){U^\dag }\\
      &=& {v_{\rm{F}}}{p_x}{\sigma _y} + {g_0}{\sigma _x} + \frac{{\hbar \omega }}{2}{\sigma _y} - {E_{Fl}}\ .
\end{eqnarray}
Since the Hamiltonian ${H_r}$ is time-independent, we can easily obtain the eigenfunctions of
$H_{r}$ as
\begin{equation}\label{eq8}
\psi _r^ \pm \left( {x,t} \right) = \frac{1}{{\sqrt 2 }}\left({\begin{array}{cccc}
\!\!\!1\!\!\! \\ \!\!\!\pm {e^{i\beta }}\!\!\!
\end{array}}\right){e^{i{k_x}x - i\left( {{\varepsilon _ \pm } - {E_{Fl}}} \right)t/\hbar }},
\end{equation}
where $e^{i\beta}=[{g_0}+i(v_{\rm{F}}\hbar k_x+\hbar\omega/2)]/\left| {\varepsilon _ \pm } \right|$. The corresponding eigenenergies are ${\varepsilon _ \pm } = \pm \sqrt {g_0^2 + {{\left( {{v_{\rm{F}}}\hbar k_x + \hbar \omega /2} \right)}^2}}$. The system has a band gap between
$-g_{0}$ and $g_{0}$ with ${g_0} > 0$. To ensure ${\psi _l}\left( {x,t} \right) = {U^{ - 1}}{\psi _r}\left( {x,t} \right)$ to be periodic in time, the eigenenergies
need to take quantized values ${\varepsilon _ \pm } - E_{Fl} = \left( {n + 1/2} \right)\hbar \omega$,
from which the allowable values of $k_x$, namely, $k_n$
for any given $E_{Fl}$ can be obtained.

We define a Floquet spin Chern number for the spin-up electrons
\begin{equation}\label{eq9}
C_F^ \uparrow = \frac{1}{\pi }\int_0^T {dt} \int_{ - \infty }^\infty  {dk_x{\mathop{\rm Im}\nolimits} \left\langle {{\partial _t}\psi _l^ - \left| {{\partial _{k_x}}\psi _l^ - } \right.} \right\rangle }\ .
\end{equation}
Equation (9) is an extension of the adiabatic spin Chern number to the more general
non-adiabatic regime.
In the adiabatic limit, $\hbar\omega\rightarrow 0$ and $E_{Fl}$ becomes negligible,
$\psi_l^{-}$ recovers the adiabatic wavefunction, and one can easily see that
the Floquet spin Chern numbers just become the adiabatic spin Chern numbers.
By substituting $\psi _l^ - \left( {x,t} \right) = {U^{ - 1}}\psi _r^ - \left( {x,t} \right)$ and equation (8) into equation (9), it is straightforward to derive  the Floquet spin Chern number to be $C_F^ \uparrow = 1$. Similarly, for the spin-down electrons, through using the inverse unitary transformation $U = {e^{i\omega t{\sigma _y}/2}}$ , one can obtain $C_F^ \downarrow =  - 1$. Therefore,
the resulting Floquet spin Chern numbers
are the same to the adiabatic ones. The system is topological,
and can pump pure spin.

\section{Spin pumping with TR symmetry}\label{s3}
\begin{figure}
\begin{center}
\includegraphics[width=0.9\textwidth]{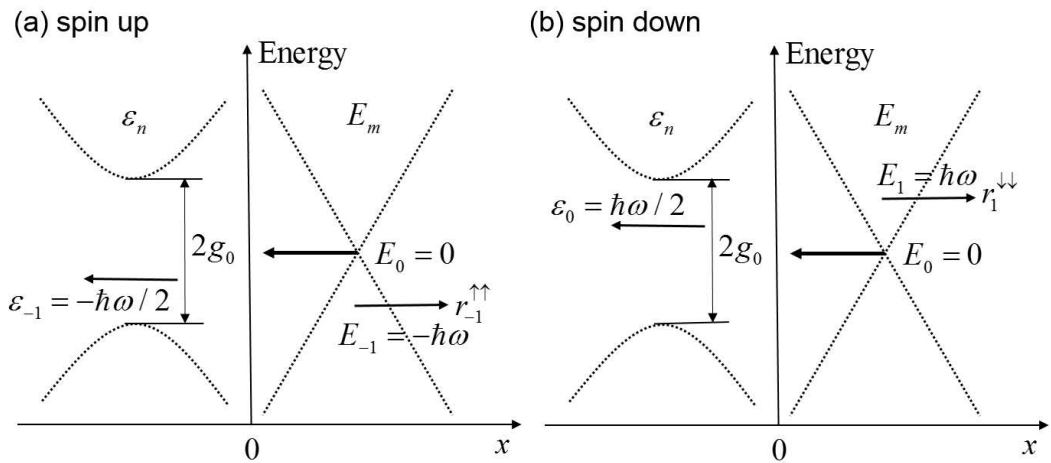}
\caption{The Floquet scattering processes. (a) For an electron wave with energy ${E_0}=E_{{\rm{F}}}=0$ incoming from the spin-up channel of the lead, the reflection wave mode has an energy ${E_{-1}} =-\hbar \omega$,
with $\hbar \omega $ as the energy of a single photon. In our model, only the single-photon processes can occur, and the multi-photon assisted processes are prohibited. In the pump part,
the transmission wave mode has an energy ${\varepsilon_{-1}} = -\hbar \omega/2$ in the rotating reference frame. The case for an electron wave incoming from the spin-down channel is 
illustrated in (b). 
The fact that the eigenenergies in the rotating reference frame take 
half-integer values in units of $\hbar\omega$ is because 
of the unitary transformation $U = {e^{ - i\omega t{\sigma _y}/2}}$, which shifts the energy levels
by $\pm\hbar\omega/2$ with respect to those in the laboratory reference frame,
as explained below Eq.\ (12). The evanescent modes in the band gap between $-g_{0}$ and $g_{0}$ play an important role in the topological pumping process. We can see that only when $2{g_0} > \hbar \omega $, the energy levels of the transmission waves for both the spin-up and spin-down cases, i.e., $\varepsilon_{0}$ and $\varepsilon_{-1}$, are in the band gap, and the spin pumping is topological.
}\label{fig1}
\end{center}
\end{figure}

Now we consider a pumping system, where a pump body described by Hamiltonian Eq.\ (1)
lies at $x < 0$, and a lead is at $x>0$ with Hamiltonian
\begin{equation}\label{eq10}
{H_L} = {v_{\rm{F}}}{p_x}{\sigma _y}\ .
\end{equation}
Here we consider the scattering process for a spin-up electron
with energy ${E_0} =E_{\rm{F}}= 0$ incident from the
lead. The reflection
waves contain different energy modes ${E_m} = m\hbar \omega $ with $m$ as an integer,
due to absorption or emission of photons.  
There is also probability for the electron to transmit into the pump body.
The Floquet eigenenergy can be chosen to be the incident energy ${E_{Fl}} = {E_{\rm{F}}} = 0$
for convenience~\cite{Li1}. We first consider the case where the spin of the incident electron is parallel to the $z$ axis. The electron spin is conserved in the scattering process, and
will be omitted for a while. On the basis $\left( \left|1 \right\rangle ,\left|- 1 \right\rangle\right)$  with the kets as the eigenstates of  ${\sigma _z}$ , the wavefunction in the lead is given by
\begin{equation}\label{eq11}
\Psi _{Fl}^L\left( {x,t} \right) = \sum\limits_{m =  - \infty }^\infty  {\left[ {\frac{{{\delta _{m,0}}}}{{\sqrt 2 }}\left( {\begin{array}{cccc}
\!\!\!1\!\!\!  \\
\!\!\!{ - i}\!\!\!
\end{array}} \right){e^{ - i{k_m}x}} + \frac{{r_m^{ \uparrow  \uparrow }}}{{\sqrt 2 }}\left( {\begin{array}{cccc}
\!\!\!1\!\!\!  \\
\!\!\!i\!\!\!
\end{array}} \right){e^{i{k_m}x}}
}
\right]{e^{ - iE_mt/\hbar}}},
\end{equation}
where ${k_m} = m\hbar\omega /{v_{\rm{F}}}$. In the pump body, depending on the electron energy is in the gap or in the band, the wave function could be evanescent modes or propagating modes. The wave function is generally written as
\begin{equation}\label{eq12}
\Psi _{Fl}^P\left( {x,t} \right) = U^{\dagger}\sum\limits_{n =  - \infty }^\infty  {{\frac{{t_n^{ \uparrow  \uparrow }}}{{\sqrt {{{\left| {a_n} \right|}^2} + {{\left| {b_n} \right|}^2}} }}\left( {\begin{array}{cccc}
\!\!\!{a_n }\!\!\! \\
\!\!\!{b_n }\!\!\!
\end{array}} \right)
}
{e^{{ - ik_n^ \uparrow x} - i\varepsilon _{n} t/\hbar}}}\ ,
\end{equation}
where $a_n  = 1 - i\hbar \left( {{v_{\rm{F}}}k_n^ \uparrow + \omega /2} \right)/{g_0}$, $b_n  = \varepsilon _{n}/g_0 $, and $\varepsilon _{n}= (n+1/2)\hbar \omega $ are the eigenenergies 
in the pump in the rotating reference 
frame, as has been obtained in Sec.\ 2. The wave vectors $k_n^ \uparrow $  are given by ${v_{\rm{F}}}\hbar k_n^ \uparrow  =  - \hbar \omega /2\pm\sqrt {\varepsilon _n^2 - g_0^2}$, where the sign $ - $ is for ${\varepsilon _n} <  - {g_0}$, and $+$ otherwise. In Eq.\ (12), the sum function represents the wave function in the pump in the 
rotating reference frame. After a unitary transformation $U^{\dagger}=e^{i\omega t\sigma_y/2}$,
the wave function goes 
back into the laboratory reference frame, so that the wave functions in
the pump and lead can be connected directly. Here, it is worth pointing out  
that while the eigenenergies $E_{m}=m\hbar\omega$ in the lead  
are taken to be integer values in units of $\hbar\omega$,
the eigenenergies $\varepsilon _{n}= (n+1/2)\hbar \omega$ in the pump 
are half-integers in the rotating reference frame.
In Eq.\ (12), the wave function in the rotating reference 
frame can be generally considered as a 
superposition of two eigenstates
of $\sigma_y$, namely, ${\chi _ + } = {\left( {1,i} \right)^T}$ and 
${\chi _ - } = {\left( {1,-i} \right)^T}$.
When acting on $\chi_{+}$ or $\chi_{-}$, $U^{\dagger}$ becomes $e^{i\hbar\omega t/2}$
or $e^{-i\hbar\omega t/2}$, which effectively shifts the energy levels by $\hbar\omega/2$
or $-\hbar\omega/2$. This fact ensures that the energy levels in the laboratory 
reference frame take integer values, so that the energy conservation law 
for the electron-photon system can be satisfied during the scattering process, as expected. 
We recall that the energy spectrum in the pump has a band gap between $-g_{0}$
and $g_{0}$. Indeed, if $\left| {{\varepsilon _n}} \right| < {g_0}$, the wave vector
$k_{n}^\uparrow$ is a complex number, corresponding to an evanescent mode. Otherwise, the wave vector is real, corresponding to a propagating mode.

The wave functions in the pump and lead are connected through the continuity equation at $x=0$
\begin{equation}\label{eq13}
\Psi _{Fl}^L\left( {x = {0^ + },t} \right) = \Psi _{Fl}^P\left( {x = {0^ - },t} \right)\ .
\end{equation}
Substitute equations (11) and (12) into equation (13), and notice that
the wave functions on both sides of equation (13) are Fourier series in time.
When two Fourier series are equal, the components of the same order in two
series must be equal. Consequently, we derive the following equations
\begin{eqnarray}\label{eq14}
\fl
\sqrt 2 {\delta _{m,0}}\left( {\begin{array}{cccc}
\!\!\!1\!\!\! \\
\!\!\!{ - i}\!\!\!
\end{array}} \right) + \sqrt 2 r_m^{ \uparrow  \uparrow }\left( {\begin{array}{cccc}
\!\!\!1\!\!\! \\
\!\!\!i\!\!\!
\end{array}} \right)
= \frac{{t_m^{ \uparrow  \uparrow }\left( {a_m  - ib_m } \right)}}{{\sqrt {{{\left| {a_m } \right|}^2} + {{\left| {b_m } \right|}^2}} }}\left( {\begin{array}{cccc}
\!\!\!1\!\!\! \\
\!\!\!i\!\!\!
\end{array}} \right) + \frac{{t_{m - 1}^{ \uparrow  \uparrow }\left( {a_{m - 1}  + ib_{m - 1} } \right)}}{{\sqrt {{{\left| {a_{m - 1} } \right|}^2} + {{\left| {b_{m - 1} } \right|}^2}} }}\left( {\begin{array}{cccc}
\!\!\!1\!\!\! \\
\!\!\!{ - i}\!\!\!
\end{array}} \right),
\end{eqnarray}
It follows from this equation that the $m$th order reflection amplitudes are coupled
to the $m$th order and $(m-1)$th order transmission amplitudes.
We notice that because $\left[ {{p_x}{\sigma _y},{H_L}} \right] = 0$,
the electron helicity ${p_x}{\sigma _y}$ is conserved. Therefore, when the incident wave in the lead
is ${\chi _ - } = {\left( {1, - i} \right)^T}$, which is an eigenstate of ${\sigma _y}$, the reflection wave in the lead should
be the other eigenstate of $\sigma_{y}$, i.e., ${\chi _ + } = {\left( {1,i} \right)^T}$. The transmission wave functions are linear combinations of
$\chi_{+}$ and $\chi_{-}$. It is easy to find $t_{m - 1}^{ \uparrow  \uparrow }=r_{m - 1}^{ \uparrow  \uparrow } = 0$ for $m \ne 0$.
The only nonzero coefficients are
\begin{equation}\label{eq15}
r_{ - 1}^{ \uparrow  \uparrow } = \frac{{a_{ - 1}  - ib_{ - 1} }}{{a_{ - 1}  + ib_{ - 1} }}\ ,
\end{equation}
\begin{equation}\label{eq16}
t_{ - 1}^{ \uparrow  \uparrow } = \frac{{\sqrt {2\left( {{{\left| {a_{ - 1} } \right|}^2} + {{\left| {b_{ - 1} } \right|}^2}} \right)} }}{{a_{ - 1}  + ib_{ - 1} }}\ .
\end{equation}

The above result indicates that only the single-photon assisted transport happens in the topological Floquet scattering process, as illustrated in Fig.\ 1(a). If $2{g_0} > \hbar \omega$, ${\varepsilon _{ - 1}} =  - \hbar \omega /2$ is in the band gap, and the wave function in the pump is evanescent mode with exponential decay. $a_{ - 1}  = 1 - i\sqrt {{{\left( {\hbar \omega /2{g_0}} \right)}^2} - 1}$ and $b_{ - 1} =  - \hbar \omega /2g_0$ are real, so ${\left| {r_{ - 1}^{ \uparrow  \uparrow }} \right|^2} = 1$. If $2{g_0} \le \hbar \omega $, there is actual transport of probability
 into the pump through the propagating mode. By defining the electron velocity ${v_{t\left( r \right)}} = \left\langle {{\psi _{t\left( r \right)}}\left| {\partial H/\partial {k_x}} \right|{\psi _{t\left( r \right)}}} \right\rangle$, and denoting
 the component of the Floquet wave functions correspondent to $t_{ - 1}^{ \uparrow  \uparrow }\left( {r_{ - 1}^{ \uparrow  \uparrow }} \right)$ as ${\psi _{t\left( r \right)}}$, it is easy to check ${\left| {r_{ - 1}^{ \uparrow  \uparrow }} \right|^2} + {\left| {t_{ - 1}^{ \uparrow  \uparrow }} \right|^2}{v_t}/{v_r} = 1$, and the conservation law of the probability current is fulfilled.
 Similarly, by considering the case, where the spin of the incident electron is antiparallel to the $z$ axis, we can obtain $r_1^{ \downarrow  \downarrow } = {\left( {r_{ - 1}^{ \uparrow  \uparrow }} \right)^ * }$ and $t_0^{ \downarrow  \downarrow } = {\left( {t_{ - 1}^{ \uparrow  \uparrow }} \right)^ * }$, and all other transmission and reflection amplitudes are zero. The scattering process
for a spin-down incident electron is illustrated in Fig.\ 1(b). 
 We notice that while the incident electrons in the two spin channels have the same energy, the reflection electrons
 have different energies, which will give rise to the pumping effect in this Floquet transport picture.

 \begin{figure}
\begin{center}
\includegraphics[width=0.7\textwidth]{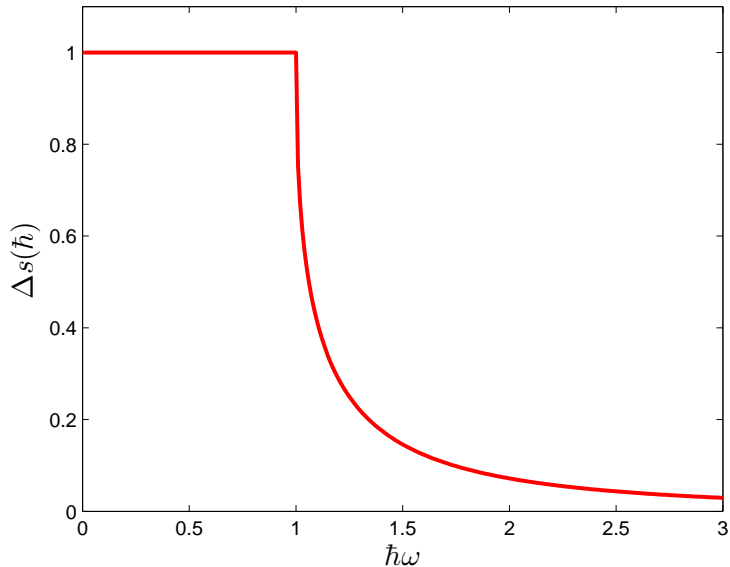}
\caption{Pumped spin $\Delta s$ as a function of the ratio of pumping frequency to the band gap. The Fermi energy is set to zero, ${E_{\rm{F}}} = 0$.
}\label{fig2}
\end{center}
\end{figure}

The spin-dependent electrical current pumped into the lead can be expressed in terms of the Floquet
scattering matrix  ${S^{s}_F}$ of the lead  as~\cite{Moskalets,Moskalets1}
\begin{equation}\label{eq17}
{I^{s}} = \frac{e}{h}\int {dE\left\{ {f_{s }^{\left( {out} \right)}\left( E \right) - f_{s }^{\left( {in} \right)}\left( E \right)} \right\}}\ .
\end{equation}
Here, $f_s^{\left( {out} \right)}\left( E \right) = \sum\limits_{n =  - \infty }^\infty  {\sum\limits_{s' =  \uparrow , \downarrow } {{{\left| {S_F^{ss'}\left( {{E_n},E} \right)} \right|}^2}f\left( {{E_n}} \right)} }$ with $s=\uparrow$ and $\downarrow$ is spin-dependent distribution function for outgoing electrons, and $f_{s }^{\left( {in} \right)}\left( E \right)$ for incoming ones,
where $f(E)$ is the Fermi distribution function. At zero temperature, substituting the scattering coefficients into equation (17), we can obtain
\begin{equation}\label{eq18}
{I^{\uparrow }} = \frac{e}{h}\int_0^{\hbar \omega } {{{\left| {r_{ - 1}^{ \uparrow  \uparrow }} \right|}^2}dE}  = \frac{{e\omega }}{{2\pi }}{\left| {r_{ - 1}^{ \uparrow  \uparrow }} \right|^2},
\end{equation}
\begin{equation}\label{eq19}
{I^{\downarrow }} = \frac{e}{h}\int_0^{-\hbar \omega } {{{\left| {r_1^{ \downarrow  \downarrow }} \right|}^2}dE}  = -\frac{{e\omega }}{{2\pi }}{\left| {r_1^{ \downarrow  \downarrow }} \right|^2}.
\end{equation}
Since $r_1^{ \downarrow  \downarrow } = {\left( {r_{ - 1}^{ \uparrow  \uparrow }} \right)^ * }$, the
pumped currents in the two spin channels are just opposite, i.e., ${I^{\uparrow }} =  - {I^{\downarrow }}$. The total charge current ${I^C} = {I^{\uparrow }} + {I^{\downarrow }}$ vanishes, and the spin current is
given by ${I^S} = \hbar \left( {{I^{\uparrow }} - {I^{\downarrow }}} \right)/2e$. It is convenient
to consider the spin pumped per cycle
$\Delta{s} = 2\pi I^{S}/\omega$, which is derived
to be
\begin{equation}\label{eq20}
\Delta s = \hbar {\left| {r_{ - 1}^{ \uparrow  \uparrow }} \right|^2} = \left\{ {\begin{array}{*{20}{c}}
\hbar &{\hbar \omega  < 2{g_0}}\\
{\frac{{1 - \lambda }}{{1 + \lambda }}\hbar }&{\hbar \omega  \ge 2{g_0}}
\end{array}} \right.,
\end{equation}
with $\lambda  = \sqrt {1 - {{\left( {2{g_0}/\hbar \omega } \right)}^2}}$.
In Fig.\ 2, the spin pumped per cycle is plotted as a function of
the single-photon energy $\hbar\omega$. When the single-photon energy
$\hbar\omega$ is less than the band gap $2g_{0}$, the pumped spin is quantized and
independent of the parameters. The spin pumping remains to be topological, which
can be attributed to the nonzero Floquet spin Chern number. While this result is consistent with the
the adiabatic approximation, we need to point out that the adiabatic approximation is valid only in the
limit $\hbar \omega \rightarrow 0$, and the present calculation is strict for any value of $\hbar\omega$.
When the single-photon energy is greater than the band gap, the pumped spin is not quantized, and dependent on the parameters. The spin
pumping is no longer related to the topology of the Floquet energy band, because the incident
electrons can now be transmitted to the Floquet propagating modes in the pump. In addition, by  similar analysis, when the Fermi energy is nonzero, either above or below the middle of the
band gap, the
region for the topological spin pumping will shrink, and the condition
for the topological spin pumping becomes $\hbar \omega /2{g_0} < 1 - \left| {{E_F}} \right|/{g_0}$.

In order to see some new features of the topological Floquet pumping, in comparison to the
topologically trivial pumping, we consider the effect of varying the strength of the
oscillating Zeeman field on the pumped spin. As can be seen from Fig.\ 2,
when the strength ${g_0}$ is large enough such that $\hbar \omega/2{g_0}<1$, the pumped spin is quantized to 2 in units of $\hbar /2$, which is the same as the adiabatic pumping~\cite{Zhou2}. However, in the topologically trivial pumping, such as the pumping in the one-dimensional system with time-dependent double delta barriers~\cite{Kim}, the Floquet pumping and the adiabatic pumping yield an equivalent result only for small strength of the oscillating Zeeman field. The difference may be understood from the fact that the strength ${g_0}$ determines the magnitude of the band gap in the topological pumping of this system.
As long as the condition $\hbar \omega  < 2{g_0}$ is fulfilled, the
pumping process is governed by the topological property of the system.

\section{Spin pumping without TR symmetry}\label{s4}
To study the spin pumping in the
absence of the TR symmetry, we include into the model a magnetic impurity
with potential ${H_M} = V(x){s_x}$. $V(x)$ is taken to be a square potential
centered at $x=0$ with height $V_{0}$ and width $d$. For simplicity, we can take the limit $d\rightarrow 0$
and keep $U_{0}=V_{0}d$ finite.  It can be shown that the scattering effect
of the impurity potential is equivalent to imposing a unitary boundary
condition for the electron wavefunctions~\cite{Zhou2}
\begin{equation}\label{eq21}
\Psi _{Fl}^L\left( {x = {0^ + },t} \right) = S\Psi _{Fl}^P\left( {x = {0^ - },t} \right),
\end{equation}
where $S = {e^{ - i\phi {s_x}{\sigma _y}}}$ with $\phi  = {U_0}/\hbar {v_{\rm{F}}}$. The
magnetic impurity explicitly breaks the TR symmetry
of the system.

We consider the scattering problem for an spin-up electron incident from the lead.
The presence of the magnetic impurity destroys the spin conservation, and we now need
to explicitly include both spin degrees of freedom in the electron wave functions.
On the basis $\left( {\left| { \uparrow ,1} \right\rangle ,\left| { \uparrow , - 1} \right\rangle ,\left| { \downarrow ,1} \right\rangle \left| { \downarrow , - 1} \right\rangle } \right)$  with the kets as the eigenstates of ${s_z}$  and ${\sigma _z}$ , the wavefunction in the lead is given by
\begin{eqnarray}\label{eq22}
\fl
\Psi _{Fl}^L = \sum\limits_{m =  - \infty }^\infty  {\left[ {\frac{{{\delta _{m,0}}}}{{\sqrt 2 }}\left( {\begin{array}{cccc}
\!\!\!1\!\!\! \\
\!\!\!{ - i}\!\!\! \\
\!\!\!0\!\!\! \\
\!\!\!0\!\!\!
\end{array}} \right){e^{ - i{k_m}x}} + \frac{{r_{M,m}^{ \uparrow  \uparrow }}}{{\sqrt 2 }}\left( {\begin{array}{cccc}
\!\!\!1\!\!\! \\
\!\!\!i\!\!\! \\
\!\!\!0\!\!\! \\
\!\!\!0\!\!\!
\end{array}} \right){e^{i{k_m}x}} + \frac{{r_{M,m}^{ \downarrow  \uparrow }}}{{\sqrt 2 }}\left( {\begin{array}{cccc}
\!\!\!0\!\!\! \\
\!\!\!0\!\!\! \\
\!\!\!1\!\!\! \\
\!\!\!i\!\!\!
\end{array}} \right){e^{i{k_m}x}}} \right]{e^{ - im\omega t}}}\ .
\end{eqnarray}
In the pump, the wave function can be written as
\begin{eqnarray}\label{eq23}
\fl
\Psi _{Fl}^P =\! \sum\limits_{n =  - \infty }^\infty  \!\left[ \frac{{t_{M,n}^{ \uparrow  \uparrow }{e^{i\omega t{\sigma _y}/2}}}}{{\sqrt {{{\left| {a_n } \right|}^2} + {{\left| {b_n } \right|}^2}} }}\!\left( {\begin{array}{cccc}
\!\!\!{a_n }\!\!\! \\
\!\!\!{b_n }\!\!\! \\
\!\!\!0\!\!\! \\
\!\!\!0\!\!\!
\end{array}} \right)\!\!{e^{ - ik_n^ \uparrow x}} +\! \frac{{t_{M,n}^{ \downarrow  \uparrow }{e^{ - i\omega t{\sigma _y}/2}}}}{{\sqrt {{{\left| {c_n } \right|}^2} + {{\left| {d_n } \right|}^2}} }}\!\left( {\begin{array}{cccc}
\!\!\!0\!\!\! \\
\!\!\!0\!\!\! \\
\!\!\!{c_n }\!\!\! \\
\!\!\! {d_n }\!\!\!
\end{array}} \right)\!\!{e^{ - ik_n^ \downarrow x}} \right]\!\!{e^{ - i\left( {n + 1/2} \right)\omega t}}\ \!\!.
\end{eqnarray}
Here, $a_n$ and $b_n$ are the same as in equation (\ref{eq12}), $c_n  = 1 - i\hbar \left( {{v_{\rm{F}}}k_n^ \downarrow - \omega /2} \right)/{g_0}$, and $d_n  = b_n $. The wave vectors $k_n^ \downarrow $  are given by ${v_{\rm{F}}}\hbar k_n^ \downarrow  =  \hbar \omega /2\pm\sqrt {{{\left[ {\left( {n + 1/2} \right)\hbar \omega } \right]}^2} - g_0^2}$,
where the sign $ - $ is for the valence band, and $+$ for the conduction band.
Similarly to solving the Floquet scattering coefficients in section~\ref{s3}, by substituting equations (\ref{eq22}) and (\ref{eq23}) into equation (\ref{eq21}) and by some algebra, we can obtain for the reflection amplitudes
$r_{M, - 1}^{ \uparrow  \uparrow } = r_{ - 1}^{ \uparrow  \uparrow }{\cos ^2}\phi$,
$r_{M,1}^{ \uparrow  \uparrow } =  - {\left( {r_{ - 1}^{ \uparrow  \uparrow }} \right)^ * }{\sin ^2}\phi$,
$r_{M, - 1}^{ \downarrow  \uparrow } =  - ir_{ - 1}^{ \uparrow  \uparrow }\sin \phi \cos \phi$,
$r_{M,1}^{ \downarrow  \uparrow } =  - {\left( {r_{M, - 1}^{ \downarrow  \uparrow }} \right)^ * }$,
where $r_{ - 1}^{ \uparrow  \uparrow }$ is the reflection amplitude in the absence of the magnetic impurity
given by equation (\ref{eq15}).
The transmission amplitudes are obtained as $t_{M,0}^{ \downarrow  \uparrow } = {\left( {it_{ - 1}^{ \uparrow  \uparrow }} \right)^ * }\sin \phi$, $t_{M, - 1}^{ \uparrow  \uparrow } = t_{ - 1}^{ \uparrow  \uparrow }\cos \phi$, where $t_{ - 1}^{ \uparrow  \uparrow }$ is given by equation (\ref{eq16}). It is easy to verify that
the reflection and transmission coefficients satisfy the conservation law of probability current. For the case of
a spin-down electron incident from the lead, the scattering amplitudes can be obtained
as $r_{M,1}^{ \downarrow  \downarrow } = {\left( {r_{M, - 1}^{ \uparrow  \uparrow }} \right)^ * }$, $r_{M, - 1}^{ \downarrow  \downarrow } = {\left( {r_{M,1}^{ \uparrow  \uparrow }} \right)^ * }$, $r_{M,1}^{ \uparrow  \downarrow } = r_{M,1}^{ \downarrow  \uparrow }$, $r_{M, - 1}^{ \uparrow  \downarrow } = r_{M, - 1}^{ \downarrow  \uparrow }$, $t_{M,0}^{ \downarrow  \downarrow } = {\left( {t_{M, - 1}^{ \uparrow  \uparrow }} \right)^ * }$, $t_{M, - 1}^{ \uparrow  \downarrow } =  - {\left( {t_{M,0}^{ \downarrow  \uparrow }} \right)^ * }$.
In both spin channels, only the single-photon assisted scattering process
contributes to the electron transport.

From equation (\ref{eq17}), the pumped current in the presence of the magnetic impurity can be obtained
as
\begin{equation}\label{eq24}
I_M^ \uparrow = \frac{{e\omega }}{{2\pi }}\left( {{{\left| {r_{M, - 1}^{ \uparrow  \uparrow }} \right|}^2} + {{\left| {r_{M, - 1}^{ \uparrow  \downarrow }} \right|}^2} - {{\left| {r_{M,1}^{ \uparrow  \uparrow }} \right|}^2} - {{\left| {r_{M,1}^{ \uparrow  \downarrow }} \right|}^2}} \right)\ ,
\end{equation}
\begin{equation}\label{eq25}
I_M^ \downarrow = -\frac{{e\omega }}{{2\pi }}\left( {{{\left| {r_{M,1}^{ \downarrow  \downarrow }} \right|}^2} + {{\left| {r_{M,1}^{ \downarrow  \uparrow }} \right|}^2} - {{\left| {r_{M, - 1}^{ \downarrow  \downarrow }} \right|}^2} - {{\left| {r_{M, - 1}^{ \downarrow  \uparrow }} \right|}^2}} \right)\ .
\end{equation}
The second and fourth terms in both equations (\ref{eq24}) and (\ref{eq25})
cancel each other, since $r_{M, - 1}^{ \uparrow  \downarrow } = r_{M, - 1}^{ \downarrow  \uparrow }$ and $r_{M,1}^{ \downarrow  \uparrow } =  - {\left( {r_{M, - 1}^{ \downarrow  \uparrow }} \right)^ * }$. Using the expressions for the other
reflection amplitudes, we find that the charge pumped per cycle is zero, and the spin pumped  $\Delta{s_M} = h \left( {{I_M^ \uparrow } - {I_M^ \downarrow }} \right)/2\omega e$ is
\begin{equation}\label{eq26}
\Delta {s_M} = \Delta s\cos 2\phi ,
\end{equation}
where $\Delta s$ is the spin pumped per cycle without the magnetic impurity given by equation (\ref{eq20}). For weak
impurity potential $\phi\ll 1$, the pumped spin can be expanded to be $\Delta {s_M} \approx \Delta s\left( {1 - 2{\phi ^2}} \right)$. For $\hbar\omega<2g_0$, $\Delta {s_M} = \hbar\left( {1 - 2{\phi ^2}} \right)$, which  recovers the result obtained in the adiabatic limit~\cite{Zhou2}. The impurity potential affects the spin pumping only in a perturbative manner, rather than destroys it immediately. This result proves that beyond the adiabatic regime, the topological spin pumping remains to be robust against TR symmetry breaking.

\section{Summary}\label{s5}
In summary, we have studied the non-adiabatic topological spin pumping based on an exactly solvable model. A Floquet spin Chern number is defined to
describe the nontrivial bulk band topology of the system. It is a generalization of the spin Chern number that was previously
introduced in the adiabatic limit to any periodically driven one-dimensional fermionic systems. The spin pumping is
topological and robust for pumping frequency smaller than the band gap, where the electron transport involves only
the Floquet evanescent modes in the pump. In this regime, the topological spin pumping is stable to TR
symmetry breaking. For pumping
frequency greater than the band gap, where the electron transport involves the propagating modes in the pump,
the pumping spin current decays rapidly and becomes irrelevant to the topological properties of the energy bands.

\section*{Acknowledgements}
This work was supported by the State Key Program for Basic Research of China under Grants No. 2015CB921202, No. 2014CB921103 (L.S.), No. 2011CB922103, and No. 2010CB923400 (D.Y.X.); the National Natural Science Foundation of China under Grants No. 11225420 (L.S.), No. 11174125, and No. 91021003 (D.Y.X.); and a project funded by the PAPD of Jiangsu Higher Education Institutions.

\renewcommand{\theequation}{A.\arabic{equation}}
\setcounter{equation}{0}


\section*{References}
\begin{thebibliography}{99}

\bibitem{Thouless} Thouless D J 1983 \emph{Phys. Rev. B} \textbf{27} 6083

\bibitem{Niu} Niu Q and Thouless D J 1984 \emph{J. Phys. A} \textbf{17} 2453

\bibitem{TKNN} Thouless D J, Kohmoto M, Nightingale M P and den Nijs M 1982 \emph{Phys. Rev. Lett.} \textbf{49} 405

\bibitem{Zhou} Zhou H Q, Cho S Y and McKenzie R H 2003 \emph{Phys. Rev. Lett.} \textbf{91} 186803

\bibitem{Qi1} Qi X L and Zhang S C 2009 \emph{Phys. Rev. B} \textbf{79} 235442

\bibitem{Mahfouzi} Mahfouzi F, Nikoli\'{c} B K, Chen S H and Chang C R 2010 \emph{Phys. Rev. B} \textbf{82} 195440

\bibitem{Ueda} Ueda H T, Takeuchi A, Tatara G and Yokoyama T 2012 \emph{Phys. Rev. B} \textbf{85} 115110

\bibitem{Wang} Wang L, Troyer M and Dai X 2013 \emph{Phys. Rev. Lett.} \textbf{111} 026802

\bibitem{Gibertini} Gibertini M, Fazio R, Polini M and Taddei F 2013 \emph{Phys. Rev. B} \textbf{88} 140508(R)

\bibitem{Marra} Marra P, Citro R and Ortix C 2015 \emph{Phys. Rev. B} \textbf{91} 125411

\bibitem{Buttiker} B\"{u}ttiker M, Thomas H and Pr\^{e}tre A 1994 \emph{Z. Phys. B} \textbf{94} 133

\bibitem{Brouwer} Brouwer P W 1998 \emph{Phys. Rev. B} \textbf{58} 10135(R)

\bibitem{Graf} Graf G M and Ortelli G 2008 \emph{Phys. Rev. B} \textbf{77} 033304

\bibitem{Braunlich} Br\"{a}unlich G, Graf G M and Ortelli G 2010 \emph{Commun. Math. Phys.} \textbf{295} 243

\bibitem{Swithes} Swithes M, Marcus C M, Campman K and Gossard A C 1999 \emph{Science} \textbf{283} 1905

\bibitem{Zhu} Zhu S L and Wang Z D 2002 \emph{Phys. Rev. B} \textbf{65} 155313

\bibitem{Moskalets} Moskalets M and B\"{u}ttiker M 2002 \emph{Phys. Rev. B} \textbf{66} 035306;
    Moskalets M and B\"{u}ttiker M 2002 \emph{Phys. Rev. B} \textbf{66} 205320

\bibitem{Kim} Kim S W 2002 \emph{Phys. Rev. B} \textbf{66} 235304

\bibitem{Prada} San-Jose P, Prada E, Kohler S and Schomerus H 2011 \emph{Phys. Rev. B} \textbf{84} 155408

\bibitem{Kane} Kane C L and Mele E J 2005 \emph{Phys. Rev. Lett.} \textbf{95} 226801

\bibitem{Bernevig} Bernevig B A and Zhang S C 2006 \emph{Phys. Rev. Lett.} \textbf{96} 106802

\bibitem{Konig} K\"{o}nig M, Wiedmann S, Brune C, Roth A, Buhmann H, Molenkamp L W, Qi X L and Zhang S C 2007 \emph{Science} \textbf{318} 766

\bibitem{Hasan} Hasan M Z and Kane C L 2010 \emph{Rev. Mod. Phys.} \textbf{82} 3045

\bibitem{Qi} Qi X L and Zhang S C 2011 \emph{Rev. Mod. Phys.} \textbf{83} 1057

\bibitem{Kane1} Kane C L and Mele E J 2005 \emph{Phys. Rev. Lett.} \textbf{95} 146802

\bibitem{Sheng} Sheng D N, Weng Z Y, Sheng L and Haldane F D M 2006 \emph{Phys. Rev. Lett.} \textbf{97} 036808

\bibitem{Prodan} Prodan E 2009 \emph{Phys. Rev. B} \textbf{80} 125327; Prodan E 2010 \emph{New J. Phys.} \textbf{12} 065003

\bibitem{Li} Li H C, Sheng L, Sheng D N and Xing D Y 2010 \emph{Phys. Rev. B} \textbf{82} 165104

\bibitem{Yang} Yang Y, Xu Z, Sheng L, Wang B G, Xing D Y and Sheng D N 2011 \emph{Phys. Rev. Lett.} \textbf{107} 066602; Yang Y, Li H C, Sheng L, Shen R, Sheng D N and Xing D Y 2013 \emph{New J. Phys.} \textbf{15} 083042

\bibitem{Sheng1} Li H C, Sheng L and Xing D Y 2012 \emph{Phys. Rev. Lett.} \textbf{108} 196806

\bibitem{Shindou} Shindou R 2005 \emph{J. Phys. Soc. Jpn.} \textbf{74} 1214

\bibitem{Fu} Fu L and Kane C L 2006 \emph{Phys. Rev. B} \textbf{74} 195312

\bibitem{Meidan} Meidan D, Micklitz T and Brouwer P W 2010 \emph{Phys. Rev. B} \textbf{82} 161303;
      Meidan D, Micklitz T and Brouwer P W 2011 \emph{Phys. Rev. B} \textbf{84} 195410

\bibitem{Sharma} Sharma P and Chamon C 2001 \emph{Phys. Rev. Lett.} \textbf{87} 096401

\bibitem{Citro} Citro R, Romeo F and Andrei N 2011 \emph{Phys. Rev. B} \textbf{84} 161301(R)

\bibitem{Keselman} Keselman A, Fu L, Stern A and Berg E 2013 \emph{Phys. Rev. Lett.} \textbf{111} 116402

\bibitem{Zhou2} Zhou C Q, Zhang Y F, Sheng L, Shen R, Sheng D N and Xing D Y 2014 \emph{Phys. Rev. B} \textbf{90} 085133

\bibitem{Chen} Chen M N, Sheng L, Shen R, Sheng D N and Xing D Y 2015 \emph{Phys. Rev. B} \textbf{91} 125117

\bibitem{Li1} Li W and Reichl L E 1999 \emph{Phys. Rev. B} \textbf{60} 15732

\bibitem{Moskalets1} Moskalets M 2012 \emph{Scattering Matrix Approach to Non-stationary Quantum Transport} (Singapore: World Scientific)

\end{thebibliography}
\end{document}